\documentclass[twocolumn,amsmath,amssymb,superscriptaddress,longbibliography,reprint]{revtex4-1}
\usepackage{graphicx}
\usepackage[dvipdfmx]{color}
\usepackage{dcolumn}
\usepackage{bm}
\usepackage{txfonts}
\usepackage[utf8]{inputenc}
\usepackage[T1]{fontenc}
\usepackage{mathptmx}
\usepackage{etoolbox}
\usepackage{multirow}
\usepackage[hypertex,colorlinks=true,linkcolor=black,citecolor=blue,filecolor=blue,urlcolor=blue,setpagesize=false,nesting=true]{hyperref}

\makeatletter
\def\@email#1#2{%
 \endgroup
 \patchcmd{\titleblock@produce}
  {\frontmatter@RRAPformat}
  {\frontmatter@RRAPformat{\produce@RRAP{*#1\href{mailto:#2}{#2}}}\frontmatter@RRAPformat}
  {}{}
}%
\makeatother
\newcommand{\msr}{$\mu$SR}
\newcommand{\IMSS}{Muon Science Laboratory and Condensed Matter Research Center, Institute of Materials Structure Science, High Energy Accelerator Research Organization (KEK-IMSS), Tsukuba, Ibaraki 305-0801, Japan}
\newcommand{\Sokendai}{Graduate Institute for Advanced Studies, SOKENDAI}
\newcommand{\NIMS}{National Institute for Materials Science (NIMS), Tsukuba, Ibaraki 305-0047, Japan}

\begin{document}
\title{Superconductivity on the verge of metal-insulator transition in Cu$_{1-x}$Zn$_x$Ir$_2$S$_4$ probed by $\mu$SR}

\author{Kenji M. Kojima}\thanks{Present address: TRIUMF, Vancouver V6T2A3, Canada}
\affiliation{\IMSS}
\author{Masanori Miyazaki}\thanks{Present address: Graduate School of Engineering, Muroran Institute of Technology, Muroran 050-8585, Japan}
\affiliation{\IMSS}
\author{Masatoshi Hiraishi}
\affiliation{\IMSS}
\author{Akihiro Koda}
\affiliation{\IMSS}\affiliation{\Sokendai}
\author{Ryosuke~Kadono}\thanks{ryosuke.kadono@kek.jp}
\affiliation{\IMSS}
\author{Christoper Baines}
\affiliation{Swiss Muon Source, Paul Scherrer Institute, 5232 Villigen PSI, Switzerland}
\author{Yoshinori Tsuchiya}
\affiliation{\NIMS}
\author{Hiroyuki S. Suzuki}\thanks{Present address: The Institute for Solid State Physics, the University of Tokyo, Kashiwa 277-0882, Japan}
\affiliation{\NIMS}
\author{Hideaki Kitazawa}
\affiliation{\NIMS}

\begin{abstract}%
The thiospinel CuIr$_2$S$_4$ undergoes a metal-insulator transition below $\approx$230 K, which is suppressed by substitution of Cu with Zn (Cu$_{1-x}$Zn$_x$Ir$_2$S$_4$) to induce superconductivity for $0.2\lesssim x\lesssim0.8$. We show that the temperature/field dependence of superfluid density in samples with $x = 0.3$ and 0.4 ($T_{\rm c} \approx 3$ K and 2.5 K) investigated by muon spin rotation and relaxation ($\mu$SR) is consistent with a fully gapped $s$-wave pairing.  Meanwhile, the relatively high resistivity (10$^{-3}$--10$^{-2}$ $\Omega\:$cm) in their normal state suggests that the superconductivity is in the ``dirty limit'' where the mean free path is much shorter than the coherence length ($\ell \ll \xi_0$). This indicates that the potential anisotropy associated with unconventional pairing mechanisms expected under the strong electron correlations is smeared out by the electron scattering.  Based on these observations, we discuss potential link between the Zn substitution-induced superconductivity and that recently discovered in CuIr$_2$S$_4$ under high pressure ($>18$ GPa) where the existence of strong electron scattering is also suggested. 
\end{abstract}
\maketitle

The thiospinel compound CuIr$_2$S$_4$ has long fascinated the condensed matter physics community due to its complex interplay between charge, orbital, and lattice degrees of freedom. At ambient pressure, this material undergoes a dramatic metal-insulator (MI) transition at $T_{\rm MI}\approx230$ K, accompanied by a structural transformation from a cubic to a triclinic phase \cite{Nagata:94,Furubayashi:94,Hagino:95}. This transition is characterized by a unique charge-ordering pattern where Ir sites separate into Ir$^{3+}$ and Ir$^{4+}$ valences, leading to the formation of isomorphic octamers through Ir$^{4+}$-Ir$^{4+}$ spin dimerization \cite{Radaelli:02}. The origin of this transition is deeply rooted in the geometric frustration inherent to the pyrochlore sublattice of Ir ions, which suppresses long-range magnetic order and instead promotes the emergence of a complex molecular-like orbital state \cite{Khomskii:05}. Understanding how this robust insulating state can be destabilized to give way to metallic or superconducting phases remains a central challenge in the study of frustrated transition metal spinels \cite{Oda:95,Oomi:95,Kumagai:95,Matsuno:97,Nagata:98,Hayashi:00,Burkov:00,Sun:01,Ishibashi:01,Ishibashi:02,Takubo:05,Yagasaki:06,Takubo:08,Bozin:11,Zhang:13,Kojima:14,Nasu:14,Ma:17,Bozin:19,Hashimoto:22}.

Chemical substitution and physical pressure are the primary tools used to probe the stability of the MI transition in CuIr$_2$S$_4$. It is well-established that the transition is suppressed by replacing Cu with Zn effectively in Cu$_{1-x}$Zn$_x$Ir$_2$S$_4$, leading to a metallic ground state that exhibits superconductivity for $0.2\lesssim x\lesssim 0.8$ with the highest transition temperature $T_{\rm c}$ of approximately 3.5 K for $x \approx 0.2$ \cite{Suzuki:99,Cao:01,Kumagai:04}. Recent landmark studies reporting the emergence of superconductivity under ultra-high pressure uncovered a complex phase diagram featuring two distinct superconducting phases, a high-$T_{\rm c}$ phase (SC-I) reaching 18.2 K above $\approx$18 GPa and a lower-$T_{\rm c}$ phase (SC-II) above $\approx$111.8 GPa \cite{Chen:26}, drawing renewed interest in the superconducting state of this compound.

Interestingly, the normal state resistivity in CuIr$_2$S$_4$ at ambient pressure is relatively high ($\rho\ge10^{-3}$ $\Omega\:$cm above $T_{\rm MI}$), and it exhibits temperature dependence characteristic to variable-range hopping \cite{Burkov:00,Cao:01,Yagasaki:06}.  The Zn substitution tends to enhance $\rho$ in the normal state by an order of magnitude \cite{Suzuki:99,Cao:01}, indicating that the system is close to the Mott-Ioffe-Regel (MIR) limit where the electronic mean free path $\ell$ is much shorter than the inverse Fermi wavevector ($k_{\rm F}\ell\lesssim1$) \cite{Ioffe:60,Mott:74,Emery:95}. From this viewpoint, the electrical resistance under ultra-high pressure suggest that CuIr$_2$S$_4$ is in the verge of MI transition over the entire pressure range up to 224 GPa. Moreover, the earlier high-pressure studies showed that the resistivity in the metallic phase up to 2 GPa approaches that of the Zn-substituted compounds at ambient pressure \cite{Oomi:95}. This similarity hints at the MIR limit as a common physical background between the superconductivity under ultra-high pressure and that induced by Zn substitution.  

In this study, we investigate the superconductivity of Cu$_{1-x}$Zn$_x$Ir$_2$S$_4$ ($x=0.3$ and $0.4$) using transverse-field muon spin rotation (TF-$\mu$SR) measurements. $\mu$SR is a unique microscopic probe that allows for the determination of the magnetic penetration depth $\lambda$ in the type II superconductors. Our results show that the temperature and magnetic field dependence of the spin relaxation rate $\sigma_{\rm s}$ ($\propto1/\lambda^2$) in both samples are consistent with a fully gapped, $s$-wave pairing symmetry. While the strong electronic correlation and the proximity to the MI transition might favor the possibility of unconventional or anisotropic pairing, the aforementioned high resistivity indicates that our samples are in the dirty limit ($\ell < \xi_0$, with $\xi_0$ being the BCS coherence length). In such a regime, the scattering by defects and/or charge fluctuations associated with the frustrated lattice is expected to wash out any gap anisotropy, effectively stabilizing an isotropic $s$-wave-like response even if the underlying pairing mechanism is non-trivial. By situating our findings within the context of the newly discovered ultra-high-pressure phases, we discuss the robustness of the superconducting state in the presence of strong scattering and its implications for the pairing mechanism in iridium-based spinels.

Polycrystalline samples of Cu$_{1-x}$Zn$_x$Ir$_2$S$_4$ used in this study were prepared using a chemical reaction in a sealed evacuated quartz ampoule. The details of sample preparation were described elsewhere \cite{Cao:01}. The bulk properties of these samples were investigated by variety of methods including magnetic susceptibility, specific heats, and powder x-ray diffraction measurements (combined in part with synchrotron radiation) to confirm the reproducibility of the earlier results.

Conventional $\mu$SR measurements were performed using the LTF spectrometer on the $\pi$M3 beamline at PSI, Switzerland and the DR spectrometer installed on the M15 beamline at TRIUMF, Canada. 
During the measurements under a transverse external field (TF), the initial muon spin direction ${\bm P}_\mu(0)$ was nearly perpendicular to the muon beam direction $\hat{z}$.  Time-dependent muon polarization was monitored by measuring decay-positron asymmetry along the $\hat{x}$ axis,
\begin{equation}
A_{x}(t) =  A_0\hat{x}\cdot{\bm P}_\mu(t)=A_0G_x(t)\cos(\omega_\mu t+\phi),\label{asy}
\end{equation}
where $A_0$ is the initial asymmetry, $G_x(t)$ is the envelop function, $\omega_\mu$ is the muon precession frequency, and $\phi$ is the initial phase of spin precession. All the measurements were performed by field-cooled process to minimize disorder of  flux line lattice (FLL) due to flux pinning. 

Applying a transverse field $B_0$ ($\parallel\hat{z}$) to sample in the superconducting state induces the formation of the FLL, bringing inhomogeneity in the local field $B({\bm r})$ along the $\hat{z}$ axis described by the density distribution $n_{\rm s}(B)$.  Since $B({\bm r})$ consists of a sum of randomly sampled internal field from FLL given by $n_{\rm s}(B)$ and random local fields from nuclear magnetic moments given by the Gaussian distribution $n_{\rm n}(B)$,  the depolarization should be provided by
\begin{eqnarray}
	\hat{x}\cdot {\bm P}_\mu(t)&=& \int_{-\infty}^{\infty}e^{-i(\gamma_{\mu}Bt)}n_{\rm s}(B-B_0-B')n_{\rm n}(B'-B_0)dBdB'\nonumber\\
	&\approx& e^{-i(\gamma_{\mu}B_0t)}\int_{-\infty}^{\infty}e^{-i(\gamma_{\mu}Bt)}n_{\rm s}(B-B')n_{\rm n}(B')dBdB',\label{gxt}
\end{eqnarray}
where $\gamma_{\mu}=2\pi\times135.53$ MHz/T is the muon gyromagnetic ratio. Considering that Eq.~(\ref{gxt}) is a Fourier transform of $n_{\rm s}(B)$ and $n_{\rm n}(B)$ in convoluted form, $G_x(t)$ is a product of the respective Fourier transforms. In the case of relatively long magnetic penetration depth ($\geq0.3$~$\mu$m) and/or of polycrystalline samples, Gaussian distribution is a reasonable approximation for $n_{\rm s}(B)$, yielding
\begin{align}
	G_x(t)\approx \exp\left[-\frac{1}{2}(\sigma_{\rm s}^2+\sigma_{\rm n}^2)t^2\right]=\exp\left[-\frac{1}{2}\sigma^2t^2\right],\label{gauss}
\end{align}
where $\sigma_{\rm s}$ and $\sigma_{\rm n}$ are the linewidths determined by the second moment of the respective field distributions, and $\sigma$ is the total linewidth to be deduced from curve fit using Eq.~(\ref{asy}). Specifically, the one associated with FLL 
is given by \cite{Brandt:88}
\begin{equation}
\sigma_{\rm s}\approx c_0\frac{\gamma_\mu\Phi_0}{\lambda^2}=c_0\frac{\gamma_\mu\Phi_0 e^2}{m^*c^2}n_{\rm s},\label{sgm_s}
\end{equation}
where $c_0=0.0607$, $\Phi_0$ ($=2.0678\times10^{-15}$ T m$^2$) is the flux quantum, $m^*$ is the effective mass of the Cooper pairs, and $n_{\rm s}$ is the superfluid density in the classical two-fluid model \cite{Gorter:34,Landau:41,Bardeen:58} (corresponding to the superconducting order parameter $|\psi^2|$ in the Ginzburg-Landau theory). Thus, Eq.~(\ref{sgm_s}) implies that $\sigma_{\rm s}\propto n_{\rm s}$. Regarding $\sigma_{\rm n}$, the experimental evaluation of nuclear dipolar fields was performed in our earlier $\mu$SR studies under zero/longitudinal field and reported elsewhere \cite{Kojima:14}, in which the detailed discussions on muon site are also found. 

\begin{figure}[t]
 \centering
  \includegraphics[width=0.90\linewidth,clip]{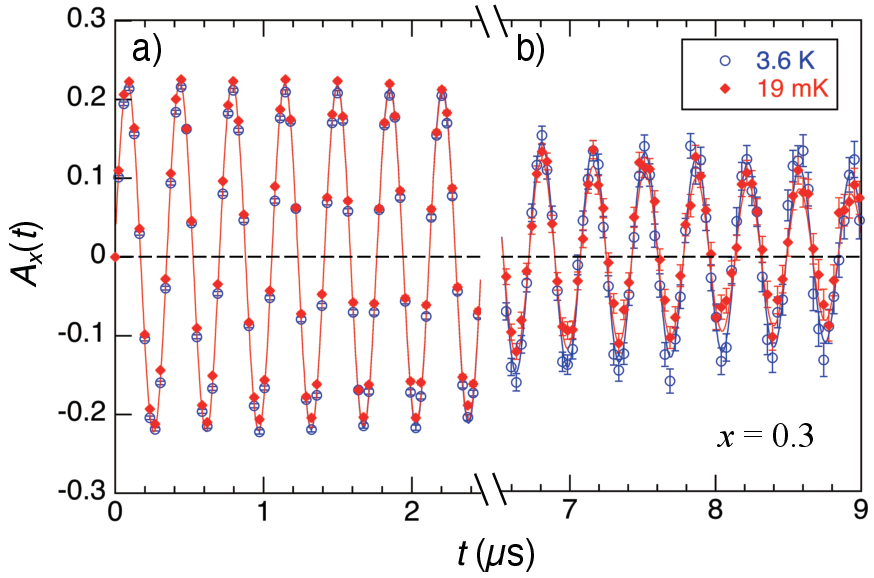}
  \caption{(Color online). \msr\ time spectra at (a) 0--2.5 $\mu$s and (b) 6.5--9 $\mu$s, observed under a transverse field (20 mT), where the spectrum measured at 19 mK (red diamonds) is overlayed with that at 3.6 K (blue circles). Solid curves are the best fits by Eqs.~(\ref{asy}) and (\ref{gauss}).}
  \label{tspec}
\end{figure}

\begin{figure}[t]
 \centering
  \includegraphics[width=0.7\linewidth,clip]{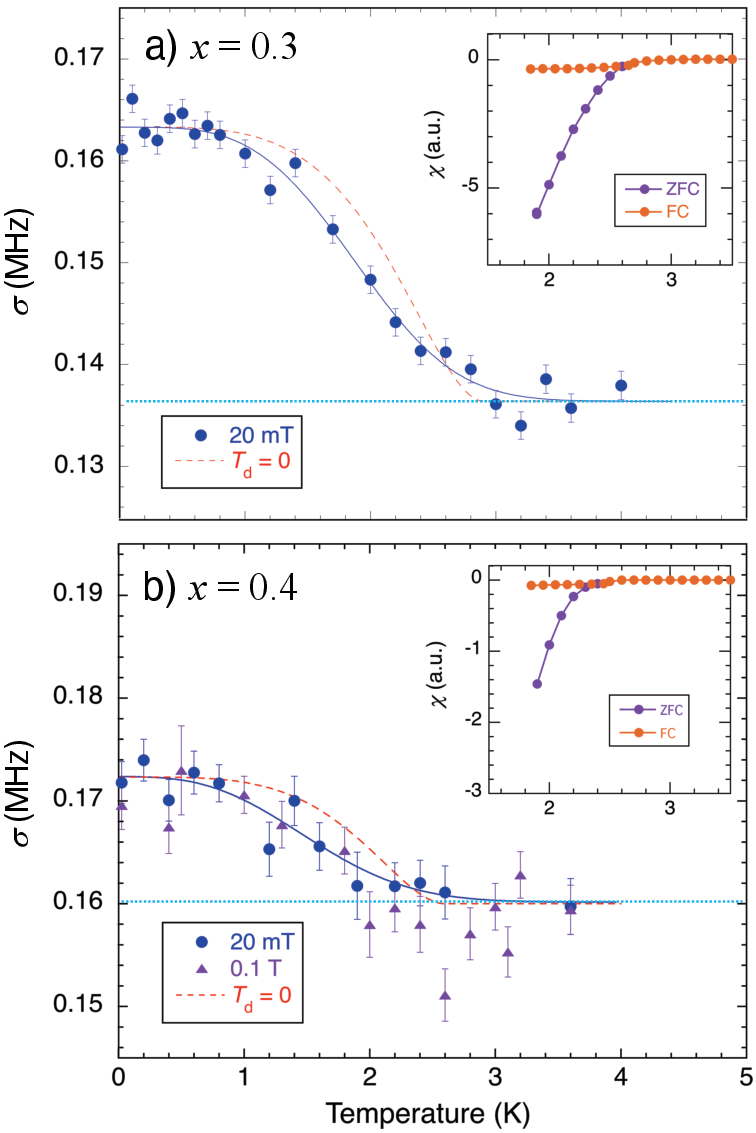}
  \caption{(Color online). Temperature dependence of the Gaussian linewidth $\sigma$ observed in the sample with Zn content $x=0.3$ (a, PSI)  and 0.4 (b, TRIUMF) . Solid line shows the result of curve fit using Eqs.~(\ref{tcdis}) and (\ref{sgmdis}), and dashed line shows expected behavior of $\sigma$ without the distribution of $T_{\rm c}$. Dotted horizontal lines correspond to $\sigma_{\rm n}$ in Eq.~(\ref{gauss}). Inset: magnetic susceptibility measured above 1.85 K under field-cool (FC) and zero-field-cool (ZFC) conditions.}
  \label{t_sigma}
\end{figure}

Figure \ref{tspec} shows typical examples of the $\mu$SR spectra observed in a sample with $x = 0.3$, divided into two time domains. In the spectrum for the later time window (6--9 $\mu$s), a decrease in asymmetry due to an increase in the relaxation rate can be observed upon the temperature drop from 3.6 K ($>T_{\rm c}$) to 19 mK ($<T_{\rm c}$). The solid lines in the figure, which represent the fit obtained using Eqs.~(\ref{asy}) and (\ref{gauss}) closely reproduces the data points. The temperature dependence of $\sigma$ obtained by the curve fits of the TF-$\mu$SR spectra in the samples with $x=0.3$ and 0.4 is shown in Fig.~\ref{t_sigma}. 

It is well established that the temperature dependence of $n_{\rm s}$ in the two-fluid model is well reproduced by the empirical equation $n_s/n\approx1-(T/T_{\rm c})^4$ (where $n$ is the total carrier density) \cite{Gorter:34}, and we have
\begin{equation}
\sigma_{\rm s}\approx\sigma_0\left[1-\left(\frac{T}{T_{\rm c}}\right)^4\right]. 
\end{equation}
Since $T_{\rm c}$ has been estimated to be 2.9(2) K at $x = 0.3$ and 2.6(1) K at $x = 0.4$ from specific heat  and resistance measurements, the temperature dependence of $\sigma_{\rm s}$ in these samples is expected to be that shown as dashed lines in Fig.~\ref{t_sigma}. While these are in line with the observed tendency at $T \lesssim0.3T_{\rm c}$ that $\sigma_{\rm s}$ is almost independent of temperature, they deviate significantly from the experimental data at $T >0.3T_{\rm c}$. Furthermore, the temperature dependence of the magnetic susceptibility (Meissner diamagnetism) shown in the inset (although the measured temperature range is limited) indicates that the development of the superconducting order parameter below $T_{\rm c}$ is rather gradual. These facts suggest that the superconducting state entails some form of inhomogeneity. Therefore, to describe such inhomogeneity, we introduce the following Gaussian distribution for $T_{\rm c}$:
\begin{equation}
P(T_{\rm c})=\frac{1}{\sqrt{2\pi}T_{\rm d}}\exp\left[-\frac{(T_{\rm c}-T_0)^2}{2T_{\rm d}^2}\right],\label{tcdis}
\end{equation}
where $T_0$ refers to the mean value of $T_{\rm c}$, and $T_{\rm d}$ to the standard deviation. The temperature dependence of $\sigma$ is then given by
\begin{equation}
\sigma^2 \approx \left\{\int_0^\infty \sigma_0\left[1-\left(\frac{T}{T_{\rm c}}\right)^4\right]P(T_{\rm c})dT_{\rm c}\right\}^2+\sigma_{\rm n}^2.\label{sgmdis}
\end{equation}
The results of fitting the data using Eq.~(\ref{sgmdis}) are shown as solid lines in Fig.~\ref{t_sigma}, and the parameter values obtained from the fits are shown in Table \ref{tab1}. For both samples, it can be seen that $T_0$ decreases to approximately $\approx$0.8--0.9$T_c$, and $T_d$ is also broad ($\approx$0.2--0.3$T_0$). As also shown in Table \ref{tab1}, the magnetic penetration depths derived from these results all exceed $\approx$1 $\mu$m. On the other hand, specific heat measurements for the sample at $x = 0.3$ indicate that the upper critical field at $T \to 0$ ($B_{c2}(0)$) is estimated to be 5.41(3) T, so the Ginzburg-Landau (GL) coherence length is estimated to be $\xi_{\rm GL}(0) = (\Phi_0/2\pi B_{\rm c2}(0))^{1/2} = 7.80(4)$ nm. Therefore, the GL parameter $\kappa=\lambda/\xi_{\rm GL}(0)$ is as large as $\approx140$, confirming that this compound is a typical type-II superconductor.

\begin{table}[t]
\begin{tabular}{cccccc}
\hline\hline
$x$ & $\sigma_0$ (MHz) & $\sigma_{\rm n}$ (MHz)& $T_0$ (K) & $T_{\rm d}$ (K) & $\lambda$ ($\mu$m) \\
\hline
0.3 & 0.090(2) & 0.136(1) & 2.64(8) K & 0.57(13) & 1.09(2)  \\
0.4 & 0.064(6) & 0.160(2) & 2.18 (30) & 0.68(46) & 1.29(4)  \\
\hline\hline
\end{tabular}
\caption{The values of the parameters ($\sigma_0$, $\sigma_{\rm n}$, $T_0$, $T_{\rm d}$) obtained by fitting the temperature dependence of the transverse relaxation rate $\sigma$ using Eq.~(\ref{sgmdis}). $\lambda$ is the magnetic penetration depth derived from $\sigma_0$.}\label{tab1}
\end{table}

\begin{figure}[t]
 \centering
  \includegraphics[width=0.75\linewidth,clip]{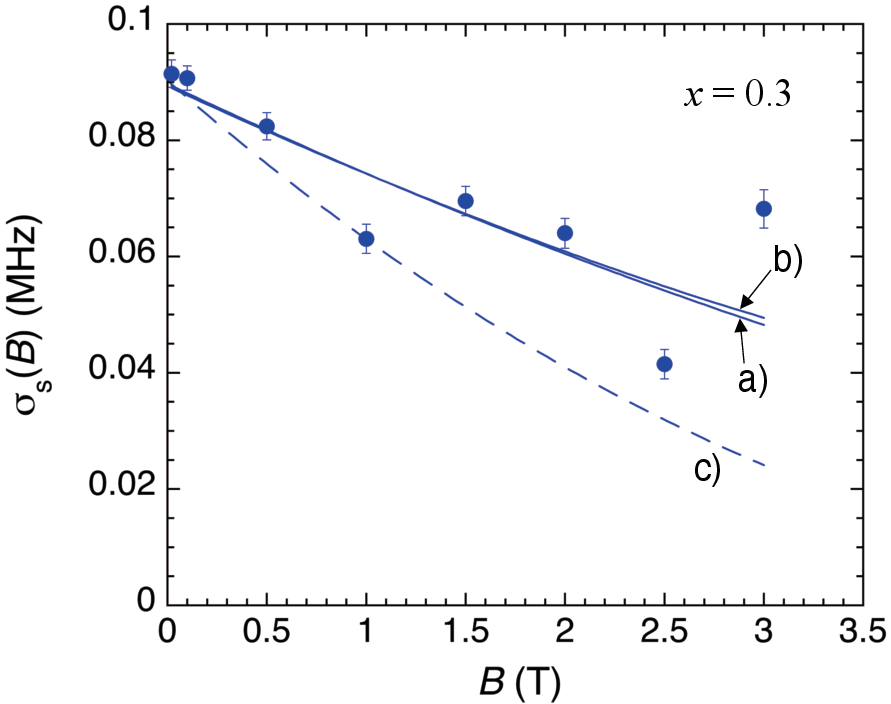}
  \caption{(Color online). $\sigma_{\rm s}$ versus $B$ obtained at 0.1 K for the sample with $x=0.3$. Solid lines show the result of curve fits (a) using Eq.~(\ref{sgmh}) with $B_{c2}(0)$ as a free parameter, or (b) using Eq.~(\ref{sgmhc}) with $\sigma_{\rm c}$ as a free parameter. Dashed line (c) shows Eq.~(\ref{sgmh}) with $B_{c2}(0)=5.41$ T.}
  \label{h_sigma}
\end{figure}

Now, let us examine the field dependence of $\sigma_{\rm s}$ for the sample with $x=0.3$ shown in  Fig.~\ref{h_sigma}, which is deduced as follows,
\begin{equation}
\sigma_{\rm s}^2(B)=\sigma^2(0.1\:\text{K};B)-\sigma^2(4\:\text{K};B),
\end{equation} 
where $\sigma(T;B)$ is the linewidth at a given temperature and field. 
The TF-$\mu$SR measurements were performed under the field-cooled condition at each field, where the stability of base temperature was $\pm0.2$ mK. Meanwhile, based on the fluctuation of  $\sigma(4\:\text{K};B)$ with respect to $B$ in the normal-conducting phase, the systematic error of $\sigma_{\rm s}(B)$ attributable to the apparatus was estimated to be $\varepsilon_B\approx0.008$ MHz.

In the FLL state, as the magnetic field approaches $B_{c2}(0)$, the overlap in the magnetic field distribution associated with the shielding current around the vortices increases, and $\sigma_{\rm s}$ is expected to approach zero due to the break down of the superconductivity. Specifically, the magnetic field dependence of $\sigma_{\rm s}$ is given by the following approximate equation \cite{Brandt:88}:
\begin{equation}
\sigma_{\rm s}(B)\approx \sigma_0f(b)=\sigma_0(1-b)\sqrt{\frac{1+3.9(1-b)^2}{4.9}},\label{sgmh}
\end{equation}
where $b=B/B_{c2}(0)$ is the normalized field, and $\sigma_{\rm s}=0$ at $b=1$.   Dashed line in Fig.~\ref{h_sigma} shows Eq.~(\ref{sgmh}) with $B_{c2}(0)$ fixed to 5.41 T.  Comparison of $\sigma_{\rm s}$ with the dashed line indicates that $\sigma_{\rm s}$ exhibits a decrease with a smaller field gradient than that predicted by Eq.~(\ref{sgmh}). 

There are two possible explanations for these results. One is that $B_{c2}(0)$ is greater than that inferred from the bulk property measurements. A curve fit using Eq~(\ref{sgmh}) with $B_{c2}(0)$ as a free parameter yields satisfactory fit with $B_{c2}(0)=10.0(6)$ T (see Fig.~\ref{h_sigma}(a)).  The other possibility is that $\sigma_{\rm s}$ includes field-independent excess relaxation $\sigma_{\rm c}$ of unknown origin.  Specifically, we assume that the magnetic field dependence of $\sigma_{\rm s}$ is given by the following equation with $B_{c2}(0)$ fixed to 5.41 T, 
\begin{equation}
\sigma_{\rm s}(B)=\sqrt{(\sigma_0^2-\sigma_{\rm c}^2)f^2(b)+ \sigma_{\rm c}^2}.\label{sgmhc}
\end{equation}
The result of curve fits using Eq.~(\ref{sgmhc}) with $\sigma_{\rm c}$ as a free parameter is shown by the solid line in Fig.~\ref{h_sigma}(b), which yields $\sigma_{\rm c}=0.036(4)$ MHz.  Although this is a non-negligible magnitude, the actual deviation of $\sigma_{\rm s}$ at 3 T from Eq.~(\ref{sgmh}) is as small as $\approx0.02$ MHz $\lesssim3\varepsilon_B$ (see Fig.~\ref{h_sigma}(b)). This invokes the possibility that such deviation is due to systematic error, and we will not discuss it further here.

In any case, the crucial point here is that there is no sign of order parameter suppression due to field-induced quasiparticle excitations (such as the Doppler shifts \cite{Volovik:93,Nakai:04}) in both cases, as would be expected in the presence of nodes and/or anisotropy in the superconducting gap. This is consistent with the conclusion derived from the temperature dependence of $\sigma_{\rm s}$. (We refer to layered nitrides as known examples where $\sigma_{\rm s}$ exhibits a sharp decrease at lower magnetic fields due to the quasiparticle excitations over a secondary gap \cite{Hiraishi:10}.)

Our TF-$\mu$SR measurements on Zn-substituted CuIr$_2$S$_4$ have revealed a fully gapped, isotropic $s$-wave superconducting state characterized by a relatively long magnetic penetration depth $\lambda \ge 1~\mu\mathrm{m}$. While the presence of strong electronic correlations and the proximity to the  MI transition might initially hint at unconventional pairing with gap anisotropy, the observed robust $s$-wave behavior can be naturally understood by situating the system deep within the ``dirty limit'' ($\ell \ll \xi_{0}$). In this regime, frequent electron scattering by structural disorder and charge fluctuations effectively averages out any potential anisotropy of the superconducting order parameter, as described by Anderson's theorem for dirty superconductors \cite{Anderson:59}. We note that such elimination of gap anisotropy was also confirmed by $\mu$SR in a borocarbide superconductor YNi$_2$B$_2$C upon substitution of Ni with Pt \cite{Ohishi:03}.

To establish a comprehensive understanding of the superconductivity in this thiospinel family, it is highly instructive to contrast the Zn-substituted system with the recently discovered superconducting phases in pristine CuIr$_2$S$_4$ under ultra-high-pressure, where a complex phase diagram featuring two distinct superconducting phases are reported \cite{Chen:26}. The emergence of the SC-I phase is driven by the successive transition from the cubic to an orthorhombic structure above $\sim$17.5 GPa that effectively suppresses an intermediate phase of dimerized triclinic insulator.  The SC-II phase emerges at 118.8 GPa, and $T_{\rm c}$ keeps increasing up to the highest pressure of 224 GPa while the SC-I phase disappears above $\approx$184 GPa around which another structural transition is inferred. These observations suggest that the pressure-induced suppression of lattice disorder effectively unmasks the intrinsic, robust superconducting potential of the Ir $5d$ electrons.

A key piece of evidence supporting the conjecture  lies in the normal-state transport properties. In pristine CuIr$_2$S$_4$ at ambient pressure, the resistivity just above $T_{\rm MI}$ is notably high ($\rho \ge 10^{-3}$ $\Omega\:$cm) \cite{Nagata:94,Furubayashi:94,Oomi:95}. At these temperatures, assuming a typical Fermi velocity $\varv_{\rm F}\approx0.5$--$1\times10^6$ m/s, the estimated $\ell\approx0.7$--1.4 nm is only a few times the Ir-Ir interatomic distance ($a/2\sim 0.5~\mathrm{nm}$), placing the parent metallic state inherently close to the Mott-Ioffe-Regel (MIR) limit (the above $\varv_{\rm F}$ corresponds to $k_{\rm F}\approx4$--9 nm$^{-1}$). When Cu is chemically substituted with Zn, the normal-state resistivity further increases by nearly an order of magnitude ($10^{-2}$ $\Omega\:$cm, suggesting $\ell\sim0.1$ nm), driving the system into the MIR limit of extreme disorder ($k_{\rm F}\ell<1$) \cite{Suzuki:99,Cao:01}. 

Crucially, a closer inspection of the high-pressure transport data in pristine CuIr$_2$S$_4$ reveals an unexpected parallel with this chemical substitution effect. Although absolute resistivity values are not explicitly given, the raw resistance in the normal state within the SC-I pressure domain appears nearly an order of magnitude higher than that observed in the metallic phase at low pressure and room temperature. This strongly suggests that the high-$T_{\rm c}$ SC-I phase shares a common physical background with the Zn-substituted compounds, operating on the verge of the MIR impurity limit due to persistent fluctuations or residual lattice anomalies. It is only when the system is driven into the ultra-high-pressure SC-II phase that the normal-state resistance drops back to a level comparable to the low-pressure room-temperature metallic phase. This implies that the SC-II phase might emerge from a cleaner framework liberated from the strong scattering regime. From this perspective, the electronic disorder ($1/\ell$) and its relation to the MIR limit serve as a primary axis for classifying the various superconducting states in this thiospinel system.

Regarding the microscopic origin of this inherent ``strong disorder'' and high resistivity in the metallic state, we suggest that it is deeply rooted in the geometrical frustration intrinsic to the spinel $B$-site (pyrochlore) sublattice formed by the Ir atoms. In CuIr$_2$S$_4$, the interplay between charge, orbital, and lattice degrees of freedom under geometrical frustration prevents the simple realization of a conventional Fermi liquid. Instead, it generates short-range fluctuations or a frustrated distribution of Ir$^{3+}$/Ir$^{4+}$ charge configurations. 

This frustration-induced fluctuation exerts a dual influence on the superconducting state. At one hand, it acts as a prominent source of strong inelastic/elastic scattering that dramatically shortens the mean free path $\ell$. This strong disorder degrades the phase coherence of Cooper pairs and suppresses the transition temperature from its intrinsic scale ($T_{\rm c}\approx18$ K) down to the observed  $\sim$3 K in the dirty limit. Meanwhile, the high density of states on the verge of the metal-insulator transition, coupled with soft phonon modes or charge/orbital fluctuations from the frustrated pyrochlore lattice, may provide a strong pairing glue. This robust pairing mechanism ensures that superconductivity survives as a fully gapped $s$-wave state without being completely destroyed by the extreme disorder.

The fact that superconductivity survives with a clean $s$-wave gap profile in Cu$_{1-x}$Zn$_x$Ir$_2$S$_4$ despite such a high normal-state resistivity demonstrates the remarkable robustness of the underlying pairing interaction. The scattering by defects and charge fluctuations washes out the gap anisotropy, protecting the superconducting state from the pair-breaking effects that usually plague unconventional superconductors in the presence of disorder.

In conclusion, our TF-$\mu$SR measurements on the thiospinel Cu$_{1-x}$Zn$_x$Ir$_2$S$_4$ ($x = 0.3, 0.4$) firmly establish an isotropic, fully gapped $s$-wave superconducting state that remains remarkably robust deep within the dirty limit ($\ell \ll \xi_{0}$). By contrasting our results with the recently reported ultra-high-pressure phases, we propose a unified framework wherein electronic disorder ($\propto1/\ell$) and proximity to the Mott-Ioffe-Regel limit serve as the primary axis governing the superconducting state. The intrinsic high-$T_{\rm c}$ potential of the Ir $5d$ electrons, which is heavily masked by strong scattering from chemical disorder and geometric frustration at ambient pressure, is systematically unmasked as the system is driven into cleaner or structurally reorganized regimes under hydrostatic pressure.

We would like to thank the PSI and TRIUMF staff for their technical support during the $\mu$SR experiments, which were conducted under the proposals 20110687 (PSI) and M1223 (TRIUMF). This work was supported by Condensed Matter Research Center, Institute of Materials Structure Science, KEK.

%
\end{document}